\documentclass[prl,amsmath,twocolumn, showpacs, superscriptaddress,10pt]{revtex4-1}
\usepackage{graphicx}
\usepackage{amsfonts}

\begin{document}

\title{Exact distributions of the number of distinct and common sites visited by $N$ independent random walkers}

\author{Anupam Kundu} 
\affiliation{Laboratoire de Physique Th\'eorique et Mod\`eles Statistiques (UMR 8626 du CNRS),
Universit\'e Paris-Sud, B\^at.\ 100, 91405 Orsay Cedex, France}
%\affiliation{ University Paris sud, Orsay, France}
\author{Satya N. Majumdar}
\affiliation{Laboratoire de Physique Th\'eorique et Mod\`eles Statistiques (UMR 8626 du CNRS),
Universit\'e Paris-Sud, B\^at.\ 100, 91405 Orsay Cedex, France}
\author{Gr\'egory Schehr}
\affiliation{Laboratoire de Physique Th\'eorique et Mod\`eles Statistiques (UMR 8626 du CNRS),
Universit\'e Paris-Sud, B\^at.\ 100, 91405 Orsay Cedex, France}

\begin{abstract}
We study the number of distinct sites $S_N(t)$ and common sites $W_N(t)$ visited by $N$ independent one dimensional 
random walkers, all starting at the origin, after $t$ time steps. We show that these two random variables can be 
mapped onto extreme value quantities associated
to $N$ independent random walkers. Using this mapping, we compute exactly their probability distributions $P_N^d(S,t)$  
and $P_N^d(W,t)$ for any value of $N$ in the limit of large time $t$, where the random walkers can be described by Brownian motions. 
In the large $N$ limit one finds that $S_N(t)/\sqrt{t} \propto 2 \sqrt{\log N} + \widetilde{s}/(2 \sqrt{\log N})$ and 
$W_N(t)/\sqrt{t} \propto \widetilde{w}/N$ 
where $\widetilde{s}$ and $\widetilde{w}$ are random variables whose probability density functions (pdfs) are computed exactly and are found to be non trivial. We verify our results through 
direct numerical simulations.
\end{abstract}

\vspace{0.5cm}
%\date{\today}

%\pacs{: }
\maketitle
%\section{Introduction}

In elementary set theory, two fundamental concepts are
the {\em union} and the {\em intersection} of a number of $N$ sets.
While the union consists of all {\em distinct} elements of the
collection of sets, the intersection consists of {\em common} elements
of all the sets. These two notions appear naturally in
everyday life: for example the area of common knowledge or the whole range of
different interests amongst the members of
a society would define respectively its stability and activity.
In an habitat of $N$ animals, the union of the territories
covered by different animals sets the geographical range of the habitat,
while the intersection refers to the common area (e. g. a water body)
frequented by all animals.

In statistical physics, these two objects are modeled respectively by the
number of distinct and common sites visited by $N$ random walkers (RWs).
The knowledge about the number of distinct sites  has applications ranging
from the annealing of defects in crystals \cite{BD,Beeler} and
relaxation processes \cite{Blu,Cze,Bor,Cond} to the spread of populations
in ecology \cite{E-K, Pie} or to the dynamics of web annotation systems~\cite{Cattuto}. Similarly the knowledge about the common area frequented by 
endangered animals is very useful for their daily health caring. Likewise, in the energy transport through a series of independent
disordered samples, the energy output will depend on the
number of energy levels common to all these materials.

Dvoretzky and Erd\"os \cite{DE} first studied the average number 
of distinct sites $\langle S_1(t) \rangle$ visited 
by a single $t$-step RW in $d$-dimensions, subsequently
studied in \cite{Vineyard,MW,FVW}. Larralde et al. generalized this to $N$ independent, 
$t$-step walkers moving on a $d$-dimensional lattice \cite{Larralde}. They found three regimes of growth (early, intermediate and late) for 
the average number of distinct sites
$\langle S_N(t) \rangle$ as a function of time. These three regimes are separated
by two $N$-dependent times scales~\cite{Larralde}. In particular they showed that in $d=1$ and $t\gg \sqrt{\log N}$, 
$\langle S_N(t) \rangle \propto \sqrt{4D~t~\log N}$ where 
$D$ is the diffusion constant of a single walker. Recently Majumdar and Tamm \cite{Majtam} studied 
the complementary quantity, namely the number of common sites $W_N(t)$ visited by $N$ walkers, each of $t$ steps, and
found analytically a 
rich asymptotic late time growth of the average $\langle W_N(t)\rangle$. They showed that in the $(N-d)$ plane there are
three distinct phases separated by two critical lines $d=2$ and $d_c(N)=2N/(N-1)$, with $\langle W_N(t)\rangle\sim t^{\nu}$
at late times where the growth exponent $\nu=d/2$ (for $d<2$), $\nu=N-d(N-1)/2$ [for $2<d<d_c(N)$] and
$\nu=0$ [for $d>d_c(N)$]~(see also~\cite{turban}). In particular,
in $d=1$, $\langle W_N(t)\rangle \sim \sqrt{4Dt}$ where the prefactor depends on $N$.
However, most of these studies were limited to the {\it average} number of distinct or common sites, and 
there exists virtually no
information about their full probability distributions, e.g. the probabilities
$P_N^d(S,t)$ that $S_N(t)=S$ and $P_N^c(W,t)$ that $W_N(t)=W$. 

Computing these distributions for general $d$-dimensional space is highly non trivial. 
Indeed, although the $N$ walkers are independent, conditioning their trajectories to a given number of 
distinct (or common) visited sites introduces strong effective correlations between them. In $d=1$, 
we show here that these random variables $S_N(t)$ and $W_N(t)$ can be mapped onto extreme values (nearest and furthest displacements) 
associated to $N$ independent walkers. This connection to extreme value statistics (EVS) allows us to compute $P_N^d(S,t)$ and $P_N^c(W,t)$ exactly for $t$ large and arbitrary $N$. We show that the induced correlations between the walkers persist 
even for $N \to \infty$ where the limiting distributions are not given by EVS of independent random variables, as erroneously 
argued in the previous study of $S_N(t)$~\cite{Larralde}.
\begin{figure}[t]
\includegraphics[width=0.9\linewidth]{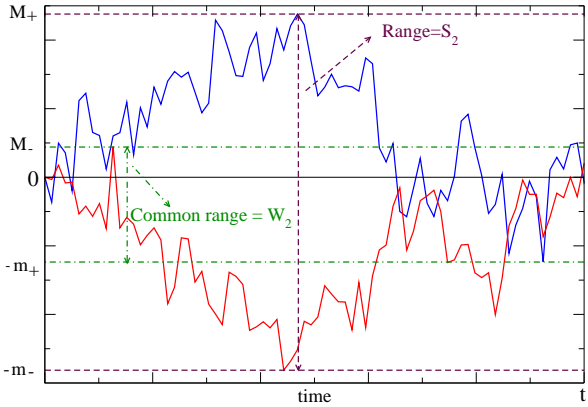}
\caption{(Color Online) Schematic diagram of $2$ independent RWs, where $M_+,~M_-,~m_+,~m_-$ and  
$S_2$, $W_2$ are shown (\ref{eq:EVS_SN}, \ref{eq:EVS_WN}). }
\label{fig1}
\end{figure}

We consider $N$ independent and identical $t$-step RWs $x_1(\tau), x_2(\tau), \cdots, x_N(\tau)$ on a $1$-$d$ lattice,
all starting at the origin. For convenience, we set the diffusion constant of the walkers $D=\frac{1}{2}$. 
Distinct sites are those that are visited at least once by at least one of the $N$ walkers
\cite{Larralde}, while common 
sites correspond to sites visited individually at least once by all the $N$ walkers~\cite{Majtam}. We denote by 
$M_i$ and $m_i$ respectively the maximum and the minimum displacements of the $i^{\rm th}$ walker $x_i$ up to time $t$.  
The number of distinct sites visited, $S_N$ \cite{foot2}, is then the sum of the range on the positive (+ve) side, $M_+$, and the range on the negative (-ve) side $m_-$
(see Fig.~\ref{fig1}):
\begin{equation}\label{eq:EVS_SN}
S_N = M_+ + m_- \,,\, M_+= \max_{1\leq i \leq N} M_i \;, \; m_-=-\min_{1 \leq i \leq N} m_i \;.
\end{equation}
Similarly, the number of common sites visited, $W_N$, is the common span on the +ve axis plus the common span $m_+$ on the -ve axis:
\begin{equation}\label{eq:EVS_WN}
W_N=M_-+m_+ \;, M_- = \min_{1\leq i \leq N} M_i \;, \; m_+ = - \max_{1 \leq i \leq N} m_i \;.
\end{equation}
Eqs. (1) and (2) establish a precise connection between $S_N$ and $W_N$ and the EVS of $N$ independent RW's.
%The relations (\ref{eq:EVS_SN}) and (\ref{eq:EVS_WN}) establish a precise and useful relation between $S_N$ and $W_N$, on the one hand, and, on the other hand, EVS of $N$ independent RW's.

In the limit of large $t$, the lattice RWs converge to Brownian motions (BMs).    
Hence for large $t$, the probability distributions $P_N^d(S,t)$ and $P_N^c(W,t)$ take the scaling form
\begin{equation}\label{eq:scaling}
P_N^{d}(S,t) = \frac{1}{\sqrt{2t}} p^{d}_N\left(\frac{S}{\sqrt{2 t}} \right) \;, \; P_N^{d}(W,t) = \frac{1}{\sqrt{2t}} p^{d}_N\left(\frac{W}{\sqrt{2 t}} \right)
\end{equation}
where $p_N^{d}(s)$ is the probability density function (pdf) of the span or range, $s = S/\sqrt{2t}$, and  $p_N^c(w)$ is the pdf of 
the common span or common range, $w = W/\sqrt{2t}$, for $N$ independent BMs (see Fig. \ref{fig1}) on the unit time interval \cite{foot1}. 
The rescaled quantities ${S_N}/{\sqrt{2t}}$ and ${W_N}/{\sqrt{2t}}$ in (\ref{eq:scaling}) are given by (\ref{eq:EVS_SN}) and (\ref{eq:EVS_WN}) where $M_\pm, m_\pm$ are replaced by their counterparts 
$\widetilde M_\pm={M_\pm}/{\sqrt{2t}}$ and $\widetilde m_\pm={m_\pm}/{\sqrt{2t}}$
corresponding to $N$ independent BMs on the unit time interval.

It is useful to summarize our main results.
We obtain exactly, for any $N$, the pdfs $p_N^d(s)$ and $p_N^c(w)$ as presented in (\ref{U-d-N-finite-1}) 
and (\ref{U-c-N-finite-1}) along with (\ref{g}) and (\ref{h}). The moments can also
be computed explicitly \cite{supp_mat}.
The tails of the pdfs can be derived explicitly:
\begin{eqnarray}\label{asympt_p_d}
p_N^d(s) \sim
\begin{cases}
&{a_N}{s^{-5}} \exp{\left[-{N\pi^2}/{(4s^2)}\right]}  \;, \; s \to 0 \;, \\
&b_N \exp{\left(-{s^2}/{2}\right)} \;, \; s \to \infty \;,
\end{cases}
\end{eqnarray}
and  
\begin{eqnarray}\label{asympt_p_c}
p_N^c(w) \sim
\begin{cases}
&c_N \, w \;, \; w \to 0 \\
&{d_N}{w^{1-N}} \exp{\left(-N\,w^2\right)} \;, \; w \to \infty \;,
\end{cases}
\end{eqnarray}
where $a_N, b_N, c_N$ and $d_N$ are computable constants (see below). For $N \to \infty$, one finds
that both pdfs approach a non trivial limiting form
\begin{eqnarray}
&&p_N^d(s) \sim 2\sqrt{\log N}~\mathcal{D}\left(2\sqrt{\log N}\left(s-2\sqrt{\log N}\right) \right) \;, \nonumber \\
&& \mathcal{D}(\widetilde{s})=2~e^{-\widetilde{s}}K_0(2~e^{-{\widetilde{s}}/{2}}) \;, \label{scale-dist}
\end{eqnarray}
where $K_n(x)$ denote the modified Bessel functions, and
\begin{eqnarray}
p_N^c(w)&=&N~\mathcal{C}\left(Nw\right),~\mathcal{C}(\widetilde{w})=\frac{4}{\pi} \, \widetilde{w} \, e^{-\frac{2}{\sqrt{\pi}}\widetilde{w}} \;, 
\; \widetilde{w} >0 \;. \;\label{scale-comm}
\end{eqnarray} 
Note that ${\cal D}(\tilde s)$ (\ref{scale-dist}) is not the Gumbel
distribution, as it was initially argued in \cite{Larralde}. 
Remarkably the same distribution ${\cal D}(\tilde s)$ also appears as the limiting distribution of the maximum of a large collection of 
logarithmically correlated random variables on a circle \cite{FB08}. We check indeed $\int_{-\infty}^{\tilde s} {\cal D}(\tilde s') d\tilde s' = 2 e^{-\tilde s/2} 
K_1(2 e^{-\tilde s/2})$, as obtained in \cite{FB08}. Incidentally, logarithmically correlated random variables have been the subject of several recent 
studies \cite{FB08,CLD01,FLDR09} because they exhibit freezing phenomena, akin to the replica symmetry breaking scenario found in mean 
field spin glass models \cite{book_sg}. As a byproduct of our computation, we show that ${\cal D}(\tilde s)$ is the convolution of two 
independent Gumbel distributions.

We start by computing the joint cumulative distribution functions (jcdf) 
$\mathbf{P}_d\left( l_1,l_2\right)$ $=\text{Pr.}\left( \widetilde{M}_+ \leq l_1,\widetilde{m}_-\leq l_2\right)$, relevant for $p_N^d(s)$ 
%(\ref{eq:EVS_SN}) 
and the jcdf $\mathbf{P}_c\left( j_1,j_2\right)$ $=\text{Pr.}\left( \widetilde{M}_- \geq j_1,\widetilde{m}_+\geq j_2\right)$ 
relevant for $p_N^c(w)$. 
%(\ref{eq:EVS_WN}). 
Since all the $N$ BMs are identical and independent, $\mathbf{P}_d\left( l_1,l_2\right)=g^N(l_1,l_2)$, where $g(l_1,l_2) = 
{\rm Pr.}(\widetilde M \leq l_1, \widetilde m \geq -l_2)$ is the jcdf of the maximum $\widetilde{M}$ and the minimum $\widetilde{m}$ for 
a {\it single} BM on the unit time interval. 
%Equivalently, $g(l_1,l_2)$ is the probability that a single Brownian motion 
%stays in the box $[-l_2,l_1]$ during a unit time interval and 
It can be computed by the standard method of images~\cite{Rednerbook}:
\begin{equation}
g(l_1,l_2)=\frac{2}{\pi} \sum_{n=0}^\infty \frac{1}{n+\frac{1}{2}}~\text{{sin}}\left(\frac{(2\,n+1)\pi l_2}{{l_1}+l_2}\right)
e^{-\left(\frac{(n+\frac{1}{2})\pi}{l_1+l_2}\right)^2} \;. \label{g}
\end{equation} 
Similarly, $\mathbf{P}_c\left( j_1,j_2\right)=h^N(j_1,j_2)$ where 
$h(j_1,j_2)=\text{Pr.}\,\left(\widetilde{M}\geq j_1, \widetilde{m}\leq -j_2 \right)$ reads: 
\begin{equation}
h(j_1,j_2)=1-\text{erf}\left(j_1\right) -\text{erf}\left(j_2\right)+g(j_1,j_2), \label{h}
\end{equation}  
where ${\rm erf}(x) = ({2}/{\sqrt{\pi}}) \int_0^x e^{-y^2} dy$, $\text{erf}\left(j_1\right)=\text{Pr.}(\widetilde{M} \leq j_1)$ and 
$\text{erf}\left(j_2\right)=\text{Prob}(\widetilde{m} \geq -j_2)$. 
From the joint pdf $\frac{\partial^2\mathbf{P}_d\left( l_1,l_2\right)}{\partial l_1\partial l_2}$ and using (\ref{eq:EVS_SN}), we obtain
\begin{equation}
p_N^d(s)= \int_0^{\infty}dl_1 \int_0^\infty dl_2~\delta(s-l_1-l_2)\frac{\partial^2g^N}{\partial l_1\partial l_2}  \;,
\label{dist-discnt-2} 
\end{equation}
with $g\equiv g(l_1,l_2)$. Similarly, from the joint pdf $\frac{\partial^2\mathbf{P}_c\left( j_1,j_2\right)}{\partial j_1\partial j_2}$ and using (\ref{eq:EVS_WN}) 
we obtain,  
\begin{equation}
p_N^c(w)= \int_0^{\infty}dj_1 \int_0^\infty dj_2~\delta(w-j_1-j_2)\frac{\partial^2h^N}{\partial j_1\partial j_2} \;, \label{dist-comm-2} 
\end{equation}
with $h \equiv h(j_1,j_2)$. For small values of $N$, the double integrals in  (\ref{dist-discnt-2}) and (\ref{dist-comm-2}) can be performed explicitly and numerical simulations confirm these exact results \cite{supp_mat} . 
Below we provide a physical interpretation of these formulas (\ref{dist-discnt-2}, \ref{dist-comm-2}) and perform, separately, their asymptotic analysis both for small and large arguments. 
We also analyze their limiting form for $N \to \infty$.
\begin{figure}[ht]
\includegraphics[width = \linewidth]{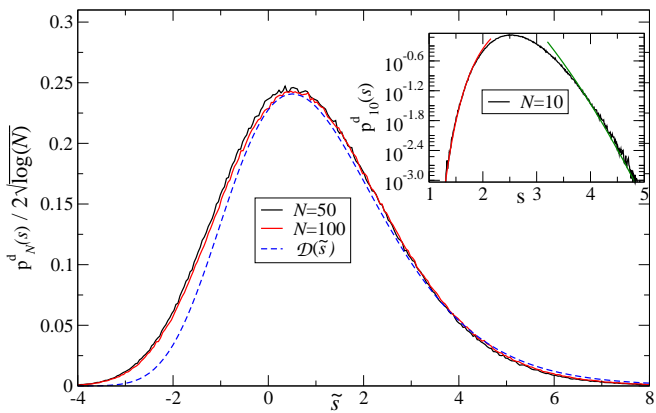}
\caption{(Color online) Plot of ${p^d_N(s)}/{(2\sqrt{\log N})}$ as a function of $\widetilde{s}=2\sqrt{\log N}\left(s- 2\sqrt{\log N}\right)$. The dotted
line indicates the exact asymptotic results for $N \to \infty$, $\mathcal{D}(\widetilde{s})$ in (\ref{scale-dist}). {\bf Inset:} Plot of 
$p^d_{10}(s)$, obtained from simulation, compared with its asymptotic behavior~(\ref{asympt_p_d}).}\label{fig:distinct}
\end{figure}

\noindent{\it{Distinct sites}} : 
To find the tails of $p_N^d(s)$ at small and large $s$ for finite $N$, we rewrite  (\ref{dist-discnt-2}) as 
\begin{eqnarray}
&&p_N^d(s)=\int_0^s dl_2 \Psi_d(s-l_2,l_2)~~\text{where}~~\label{U-d-N-finite-1} \\
 &&\Psi_d(l_1,l_2)=N~g^{N-1}\frac{\partial^2g}{\partial l_1 \partial l_2}  
+ N(N-1)g^{N-2}\frac{\partial g}{\partial l_1}\frac{\partial g}{ \partial l_2} \;. \nonumber 
\end{eqnarray}
We interpret the two contributions in $\Psi_d(l_1,l_2)$ as follows~\cite{supp_mat}: the first term 
corresponds to a configuration where  
one particle explores a region $[-l_2,s-l_2]$ (we call it a box) of size $s$ in unit time interval, 
such that its maximum is at $s-l_2$ and minimum is at
$-l_2$, while all the other $(N-1)$ 
particles stay inside this box. On the other hand, the second term corresponds to a 
configuration where two particles create, in a different way, the same box $[-l_2,s-l_2]$ of size $s$: one of the 
two particles has its maximum at $s-l_2$ and minimum larger than $-l_2$ while the second particle has its minimum at $-l_2$ 
and maximum below $s-l_2$ and all other $(N-2)$ particles stay strictly inside this box. 

When $s \to 0$ in   (\ref{U-d-N-finite-1}), 
one can replace $g(l_1,l_2)$ (\ref{g}) by its asymptotic behavior when $l_1, l_2 \to 0$ where $g(l_1,l_2)\sim \frac{4}{\pi}\text{{sin}}\left(\frac{\pi~l_2}{l_1+l_2}\right) e^{-\frac{\pi^2}{4(l_1+l_2)^2}}$. Inserting it in (\ref{U-d-N-finite-1}), we see that both terms in (\ref{U-d-N-finite-1}) contribute equally. After integration over $l_2$, one then obtains 
%decrease as $N(N-1)\frac{4\pi^2}{s^6}e^{-\frac{N\pi^2}{4s^2}}$. 
%Hence adding both terms and integrating over $l_2$ one obtains 
the result announced in~(\ref{asympt_p_d}) for $s \to 0$ 
with $a_N = 4 \pi^{3/2} N(N-1)~\left(\frac{4}{\pi}\right)^{N-2}~\frac{\Gamma(\frac{N-1}{2})}{\Gamma(\frac{N}{2})}$, where $\Gamma(x)$ 
is the Gamma function. To perform the large $s$ asymptotic of $p_N^d(s)$ we use the 
Poisson summation formula: $g(l_1,l_2)=\sum_{m=0}^{\infty} (-1)^m \left[ \text{erf}\left[m(l_1+l_2)+l_1\right] + \text{erf}\left[m(l_1+l_2)+l_2\right] \right]$. 
We use this form to evaluate the integrand in  (\ref{U-d-N-finite-1}) in the limit $s \to \infty$. 
We see that the first term in (\ref{U-d-N-finite-1}), which corresponds to create a box 
$[-l_2,s-l_2]$ with one particle, decreases as $e^{-(s+l_2)^2}e^{-l_2^2}$ whereas the second term where the same box is created
by two particles decreases as $e^{-(s-l_2)^2}e^{-l_2^2}$. Since $l_2$ is always +ve, 
the two particles term wins over the one particle term when $s \to \infty$: this is physically understandable because creating a very large span with two particles is 
more likely than creating the same one with a single particle. It also follows from this analysis that the integral over $l_2$ in (\ref{U-d-N-finite-1}) is dominated
by $l_2 \sim {\cal O}(s)$, which yields finally the large $s$ behavior announced in   (\ref{asympt_p_d}) with $b_N = {2N(N-1)}/{\sqrt{\pi}}$. 
In Fig.~\ref{fig:distinct} we verify that the small and large $s$ asymptotics of $p_N^d(s)$ given in (\ref{asympt_p_d}), for $N=10$, 
describe very well, without any fitting parameter, the distribution obtained from direct simulation, without any fitting parameter.

What happens for large $N$ ? The typical scale of the  fluctuations of $S_N/\sqrt{2 t}$ can be estimated from the relations with 
EVS (\ref{eq:EVS_SN}). The variables $\widetilde{M}_i$'s, with $i=1,\cdots,N$, which are the maxima of the $i^{\rm th}$ 
BM on the unit interval, are i.i.d. variables. Their common pdf is known to be a half-Gaussian, $p(M)=({2}/{\sqrt{\pi}}) e^{-M^2}, M>0$.
The same holds for the variables $-\widetilde{m}_i$'s. Hence, for large $N$, standard results of EVS \cite{Gumbel} state that the typical value of 
$\widetilde M_+ =\max_{1\leq i \leq N} \widetilde M_i$ is   
${\cal O}(\sqrt{\log N})$ while its fluctuations are of order $1/\sqrt{\log N}$ and governed by a Gumbel distribution. 
The same also holds for $\widetilde m_- = -\min_{1\leq i \leq N} \widetilde m_i$. For large $N$, these two extremes  
become uncorrelated as the global maximum and global minimum are most likely reached by two independent walkers. Hence one gets
\begin{equation}
g^N\left[\mu_N+\tfrac{\tilde{l}_1}{2 \mu_N},\mu_N+\tfrac{\tilde{l}_2}{2\mu_N}\right] 
\underset{N\to +\infty}{\longrightarrow} e^{-e^{-\widetilde{l}_1}}e^{-e^{-\tilde{l}_2}}\label{limit_gN}
\end{equation}
with $\mu_N = \sqrt{\log N}$. Inserting (\ref{limit_gN}) in (\ref{dist-discnt-2}) with $\widetilde{s}=2\mu_N(s-2\mu_N)$ one finds
\begin{equation}
p_N^d(s) \sim 2\sqrt{\log N}\int_{-\infty}^{\infty} d\widetilde{l}_2~e^{-\widetilde{s}}e^{-e^{-\widetilde{l}_2}}
e^{-e^{-(\widetilde{s}-\widetilde{l}_2)}} \;, \label{limit-dist-dist}
\end{equation}
which can be evaluated explicitly to give (\ref{scale-dist}). In Fig.~\ref{fig:distinct} we plot 
${p_N^d(s)}/{2\sqrt{\log N}}$ against $\widetilde{s}$ for $N=50$ and $100$. They show a relatively good agreement with the exact result 
 $\mathcal{D}(\tilde s)$ after an 
overall shift of order $\mathcal{O}({1}/{\log N})$ along the $x$-axis, thus revealing, as expected, a slow convergence towards the asymptotic result. In \cite{Larralde} the authors argued that the limiting distribution should be 
a Gumbel distribution, overlooking the fact that it is actually the {\it convolution} of two Gumbel distributions, as in (\ref{limit-dist-dist}).
In particular, for large $\tilde s$, ${\cal D}(\tilde s) \sim \tilde s e^{-\tilde s}$, while the Gumbel distribution decays as a pure exponential.   
\begin{figure}[ht]
\includegraphics[width = \linewidth]{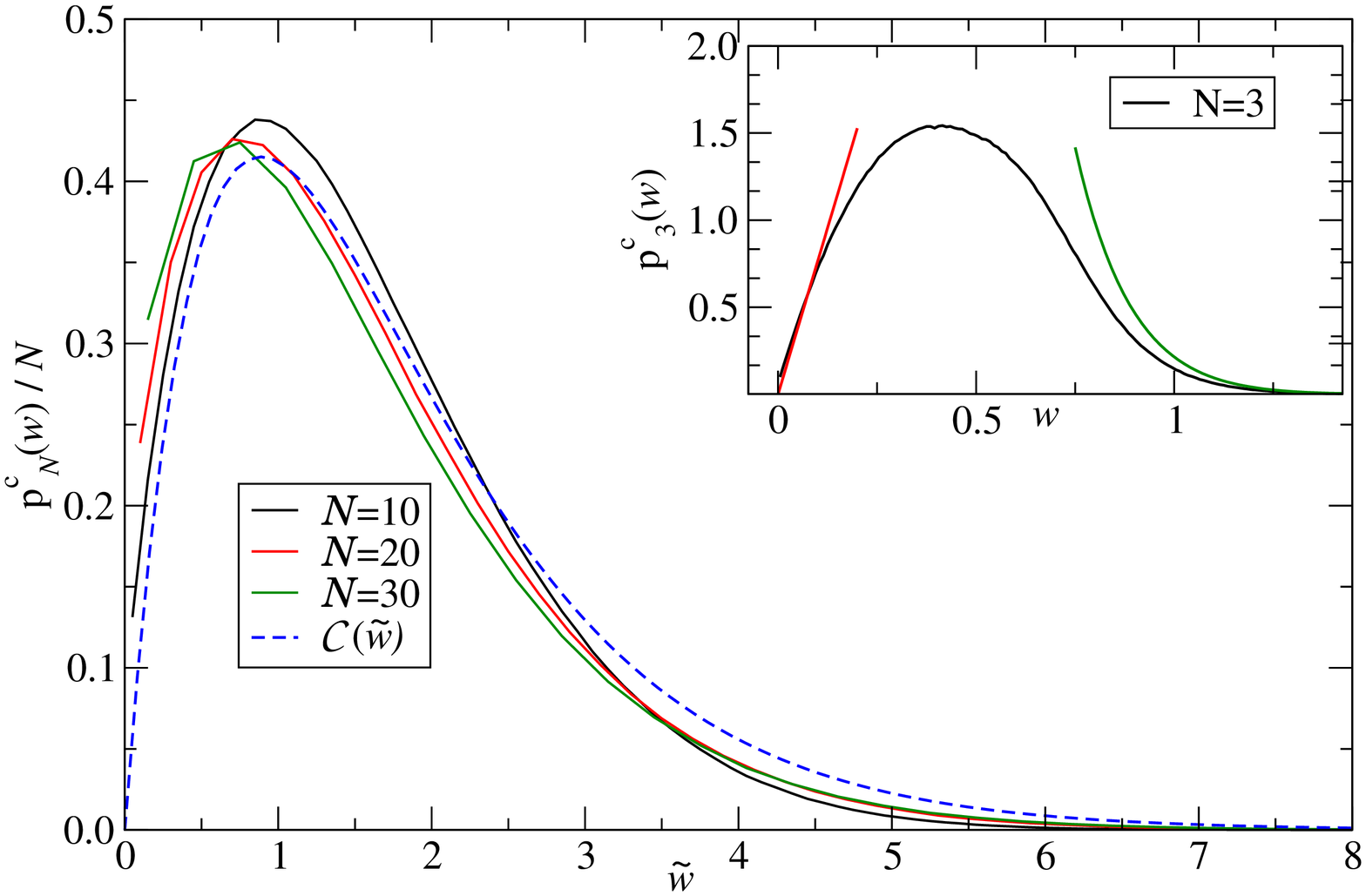}
\caption{(Color online) Plot of ${p^c_N(w)}/{N}$ as a function of $\widetilde{w}=Nw$. The dotted
line indicates the exact asymptotic results for $N \to \infty$, $\mathcal{C}(\widetilde{w})$ in (\ref{scale-comm}). {\bf Inset:} Plot of 
$p^d_{3}(w)$, obtained from simulation, compared with its asymptotic behavior~(\ref{asympt_p_c}).}\label{fig:common}
\end{figure}

\noindent {\it{Common sites}} : 
To find the small and large $w$ asymptotics of $P_N^c(w)$ we write (\ref{dist-comm-2}) as
\begin{eqnarray}
&&p_N^c(w)=\int_0^w dj_2 \Psi_c(w-j_2,j_2)~~\text{where}~~\label{U-c-N-finite-1} \\
&&\Psi_c(j_1,j_2)=N~h^{N-1}\frac{\partial^2h}{\partial j_1 \partial j_2}
+ N(N-1)h^{N-2}\frac{\partial h}{\partial j_1}\frac{\partial h}{ \partial j_2} \;. \nonumber 
\end{eqnarray}
In (\ref{U-c-N-finite-1}), one interprets the first term as one single particle creating a common span $[-j_2,w-j_2]$ of size $w$ 
and the second term as two particles collaboratively creating the same common span (in a unit time interval)~\cite{supp_mat}. In both cases, 
the remaining particles are such that their
maxima are above $w-j_2$ and their minima are below $-j_2$. When $w \to 0$ in   (\ref{U-c-N-finite-1}), $h(j_1,j_2)$ 
can be replaced by its asymptotic behavior for small $j_1, j_2$:  $h(j_1,j_2)\sim \left( 1-\frac{2}{\sqrt{\pi}}(j_1+j_2)\right)$. 
Integrating then over $j_2$ in (\ref{U-c-N-finite-1}) yields the small $w$ behavior in 
  (\ref{asympt_p_c}) with $c_N = {4N(N-1)}/{\pi}$.
Note that for very small $w$, it is much more likely to create a box of size smaller than $w$ with {\em two} particles (which occurs with a probability $\propto w^2$) than with a single one [which occurs with probability $\propto \text{exp}\left(-{\pi^2}/{4w^2}\right)$]. The former configurations thus dominate for small $w$.

To get the large $w$ behavior of $p_N^c(w)$, we estimate $h(j_1,j_2)$ for large $j_1$ (\ref{U-c-N-finite-1}). This is conveniently done by
using the Poisson formula, which yields $h(j_1,j_2) \sim \text{erfc}\left(2j_1+j_2\right) + \text{erfc}\left(j_1+2j_2\right)$. 
This estimate shows that for 
$w \gg \sqrt{\log N}$, the second term in (\ref{U-c-N-finite-1}) becomes subdominant compared to the first one. Hence for very large $w$ 
the leading  contribution comes from the first term where we replace 
$h^{(N-1)}(w-j_2,j_2) \sim [{\rm erfc}(w+j_2)+{\rm erfc}(2w-j_2)]^{N-1}$ by $\text{erfc}^{(N-1)}(w)$ as one can show that the integral over $j_2$ in (\ref{U-c-N-finite-1}) is dominated by the vicinity of $j_2 = 0$ \cite{supp_mat}. This leads to the large $w$ behavior in (\ref{asympt_p_c}) with $d_N = 8 N/\pi^{N/2}$. The asymptotic behaviors of $p_N^c(w)$ (\ref{asympt_p_c}) have been verified numerically for $N = 3$ in Fig.~\ref{fig:common}.

To obtain the typical scale of $W_N/\sqrt{2t}$ for large $N$, we use its relation to EVS~(\ref{eq:EVS_WN}). From standard EVS for i.i.d. random variables \cite{Gumbel}, we know that $\widetilde M_- = \min_{1 \leq i \leq N} M_i$, where
 $M_i \geq 0$ and distributed according to a half-Gaussian, 
is of order ${\cal O}(N^{-1})$. Its pdf is given by a Weibull law, which is here an exponential distribution \cite{Gumbel}. Indeed one has here ${\rm Pr.}(N \widetilde M_- \geq x) = e^{-\frac{2}{\sqrt{\pi}} x}$, $x>0$, as $N \to \infty$. The same holds for $\widetilde m_+$, which for large $N$ becomes independent of $\widetilde M_-$ as both of them are reached by two independent walkers. Hence, from (\ref{eq:EVS_WN}), $N W_N/\sqrt{2t}$ is given by the convolution of two exponential laws: 
\begin{equation}
p_N^c(w) \sim N^2 ({4}/{\pi})e^{-\frac{2}{\sqrt{\pi}}Nw}\int_0^{w}~dk \sim N~\mathcal{C}(Nw) \;, \label{limit-dist-comm}
\end{equation}
with $\mathcal{C}(\widetilde{w})$ as announced in   (\ref{scale-comm}). We have also obtained this result~\cite{supp_mat} by a direct large $N$ expansion of (\ref{U-c-N-finite-1}).  
In Fig.~\ref{fig:common} we plot 
${p_N^c(w)}/N$ against $\widetilde{w}$ for $N=10,~20$ and $30$ and see that they both coincide with the function~$\mathcal{C}(\tilde w)$, although the convergence is rather slow.

\noindent {\it{Conclusion}} :  We have achieved a complete analytic description of the pdfs of the number of distinct and common sites visited 
by $N$ independent RWs after $t$ time steps, for large $t$. 
We have also obtained interesting limiting distributions (\ref{scale-dist}, \ref{scale-comm}) in the limit when $N \to \infty$. 
For distinct sites, we found an intriguing connection with the maximum of logarithmically 
correlated random variables on a circle~\cite{FB08}. 

One may wonder about the effects of interactions between the walkers. For instance,
one can study non-intersecting (vicious) RWs~\cite{Fis84}. An interesting situation is 
the case where all $N$ walkers start and end at the same point, while staying positive in the time interval $[0,t]$ 
(watermelons with a wall). In this case, the number of distinct sites $S_N/\sqrt{2t}$ corresponds to the 
maximal height of these watermelons~\cite{SMCRF08}. For large $N$, the pdf of $S_N/\sqrt{2t} \propto \sqrt{N}$ properly shifted and scaled, converges to the 
Tracy-Widom distribution ${\cal F}_1$ \cite{FMS11}, which describes the fluctuations of the largest eigenvalue of 
Gaussian orthogonal random matrices. On the other hand, the number of common sites $W_N/\sqrt{2t}$ is related to 
the maximum of the lower path, the distribution of which is a very interesting open problem \cite{TW07}.

\acknowledgments
We thank A. Perret for a useful discussion. This research was supported by ANR grant 2011-BS04-013-01 WALKMAT
and in part by the Indo-French Centre for the Promotion of Advanced Research under Project $4604-3$.


\begin{thebibliography}{10}
\bibitem{foot2}
From now on, we drop the explicit $t$ dependence on $S_N, W_N, M_{i}, m_{i}$.
\end{thebibliography}
\end{document}